\documentclass[useAMS,usenatbib,referee]{mn2e}
\usepackage{epsfig}
\usepackage{graphicx}

\begin{document}
\voffset=-0.5 in

\title[Soft gamma repeaters outside the Local Group]
{Soft gamma repeaters outside the Local Group}

\author[S.B. Popov, B.E. Stern]{S.B. Popov,$^{1}$  and B.E. Stern$^{2,3,4}$
\thanks{E-mail: polar@sai.msu.ru (SBP); stern@bes.asc.rssi.ru (BES)}\\
$^1${\sl Sternberg Astronomical Institute, Moscow, 119992,
Russia} \\
$^2${\sl Institute for Nuclear Research, Pr 60-Letiya Oktyabrya, 7a, Moscow 117312, Russia }\\
$^3${\sl Astro Space Center of Lebedev Physical Institute, Profsoyuznaya 84/32,
117997, Moscow 117997, Russia }\\ 
$^4${\sl Astronomy Division, P.O.Box 3000, 90014 University of Oulu, Finland}\\ }
\date{Accepted ......  Received ......; in original form ......
      }

\maketitle

\begin{abstract}
  We propose that the best sites to search for Soft Gamma Repeaters
  (SGRs) outside the Local Group are the galaxies with active
  massive-star formation. Different possibilities to observe SGR
  activity from these sites are discussed. In particular, we have
  searched for giant flares from the nearby galaxies ($\sim~2$~--~4~Mpc away) 
  M82, M83, NGC 253, and NGC 4945 in the BATSE data.
  No candidate giant SGR flares were found. The absence of such
  detections implies that the rate of giant flares with energy release
  in the initial spike above $0.5 \cdot 10^{44}$ erg is less then 1/30
  yr$^{-1}$ in our Galaxy.  However, hyperflares similar to the one of
  December 27, 2004 can be observed from larger distances.
  Nevertheless, we do not see any significant excess of short GRBs
  from the Virgo galaxy cluster as well as from the galaxies Arp 299
  and NGC 3256 (both with extremely high star formation rate). This
  implies that the Galactic rate of hyperflares with energy release
  $\sim 10^{46}$ erg is less than $\sim 10^{-3}$ yr$^{-1}$. With this
  constraint the fraction of possible extragalactic SGR hyperflares
  among BATSE's short GRBs should not exceed few percents.  We present
  the list of short GRBs coincident with the galaxies mentioned above,
  and discuss the possibility that some of them are SGR giant flares.
  We propose that the best target for the observations of
  extragalactic SGR flares with {\it Swift} is the Virgo cluster.

\end{abstract}


\section{Introduction}

Soft gamma repeaters (SGRs) are one of the most puzzling types of
neutron stars (NS). At present, at least four of them are known in our
Galaxy and in the Large Magellanic Cloud (further, we will refer to all of
them, including the ones in the LMC, as ``Galactic'' ) 
and there are also two candidates.\footnote{Here and below we refer to \citet{wt2004}
  for the recent summary of all properties of SGRs.}

SGRs show three main types of bursts (however, these types form a
continuous spectrum of transient behavior):
\begin{itemize}
\item weak bursts, $L\la 10^{41}$~erg~s$^{-1}$;
\item intermediate bursts, $L\sim10^{41}$--$10^{43}$~erg~s$^{-1}$;
\item giant flares, $L\la 10^{45}$~erg~s$^{-1}$.
\end{itemize}

The weak bursts are relatively frequent. About several hundreds have
been detected from 4 sources during $\sim 25$ yrs, i.e. the average
rate is a few per month per source (for example, \citet{cetal1996}
report observations of 111 bursts from SGR 1806-20 during $\sim$ 5
years, see also \citep{getal2000} where the authors presented hundreds
of weak bursts and performed detailed simulations of their reccurence
time).  However, these bursts appear in groups during the periods of
activity of a SGR, so the rate is higher during these periods and
lower between. The duration of a burst is very short, $<1$ s.

The intermediate bursts have typical durations of $\sim$ few seconds
and are much more rare.  The extremely energetic giant flares (GF) are
very rare --- only three (or four, as suggested, for example, by
\citet{metal2004}\footnote{Many authors do not include the burst of
  SGR 1627-41 on June 18, 1998 as it was slightly dimmer than others
  and had no pulsating tail.  That is why it is often claimed that
  only three GFs have been detected and we accept this value in the
  rest of the paper.}) have been observed.  These bursts have a very
intensive initial spike with the duration of a fraction of a second
and a pulsating tail with a significant energy fluence but with a much
lower intensity (for the HF of SGR 1806-20 the energy emitted in the
spike was much higher than the energy in the tail).  Further, we
consider only the initial spikes as they can be confused with the
short $\gamma$~-ray bursts.  The rate of GFs is very uncertain due to
the lack of detections, usually it is estimated to be about (1/25 -
1/50) yrs$^{-1}$ per source \citep{wt2004}.

The latest GF was observed on December 27, 2004 \citep{betal2005}.
There has been a number of papers analyzing this burst
\citep{h2005,metal2005,mer2005,petal2005}.  The burst energy release
is above $10^{46}$ erg (if the distance estimate of 15 kpc is correct,
see discussion in \citep{cetal2005,mg2005}).  It is two orders of
magnitude higher than the energy release of the other GFs. It has been
suggested to be a representative of the forth class of bursts --
``supergiant flares'' or ``hyperflares'' (HFs). In principle, this
burst could form a continuous distribution together with the other
GFs. However, the huge difference in the luminosity is the reason to
consider this kind of events separately and use the term ``hyper''.


As well as being very interesting SGRs are also very rare, probably
due to their short life cycle, $\sim 10^4$~yrs.  It is suggested that
about 10\% of all NSs had been born as magnetars \citep{ketal1998} and
appeared as SGRs in their youth.  It would be very important to detect
these sources outside the Local Group.  Especially, it is interesting
to understand the birth rate of SGRs and the fraction of the NSs that
produce these sources.

Here, we would like to discuss the possibility of observing SGRs
outside the Local Group (for the previous discussions of extragalactic
SGRs see \cite{metal1982}, \citet{d2001} and the recent e-print by
\citet{ngpf2005}).  The detection of such objects will give us an
opportunity to study the properties of SGRs with larger statistics.
In this short note we focus mainly on the regions of active star
formation.  The connection between SGRs and star formation is obvious.
Being very young objects SGRs have to trace the regions of massive
star production.  The higher the star formation rate (SFR) the larger
is the number of SGRs.  In addition, recently, several authors
suggested that magnetars should be born from the most massive stars
that still produce NSs \citep{fetal2005, getal2005}\footnote{Note,
  that the NSs originated from massive progenitors are expected to be
  massive themselves \citep{whw2002}.  There are many properties, that
  distinguish massive NSs. Here we want to mention the possibility of
  solid core formation \citep{ah1983} which can lead to an opportunity
  to support strong glitches.}.
Thus, there is a clear relation between the SGR formation rate and
the star formation rate of massive stars (and so the supernova rate).

We will discuss three types of sites for the observations of
extragalactic GFs and HFs:

\begin{itemize}
\item Close-by ($<$5 Mpc) galaxies with high star formation rate
  should give the main contribution to the detection of GFs and HFs.
\item Few galaxies with extreme values of star formation rate (so-called
``supernova factories'') are the best sights to search for rare HFs.
\item HFs also can be expected to be 
detectable from the Virgo cluster of galaxies.

\end{itemize}

In the next section, we focus on the first topic, the rest two are
discussed in the third and the forth sections. We also discuss the use
of the BATSE data\footnote{http://cossc.gsfc.nasa.gov/batse/} as an
archive to search for GFs and HFs. So far, this experiment provided
the best possibilities for detection of bursts of high-energy
radiation due to its half-sky exposure, long observation time and high
sensitivity.



\section{Giant flares from nearby galaxies}

As it is discussed by \citet{h1998}, inside the 10 Mpc radius, 25\% of
star formation is due to just four well-known galaxies: M82
(d$=3.4$~Mpc), NGC 253 (2.5~Mpc), NGC 4945 (3.7~Mpc), and M83
(3.7~Mpc). Obviously, inside $\sim 4$~--~5~Mpc (this is the limiting
distance for the BATSE detection of a GF, see Fig.~3 below and the
discussion in the text) their contribution is even higher. The main
idea which we put forward here is the following: in BATSE data, the
close-by galaxies with a high present-day star formation rate are the
best sites to search for SGRs outside the Local Group.



We scale the SGR activity by the rate of supernova bursts assuming
that the number of SGRs is proportional to the supernova rate and the
activity of each source is identical.  Usually, uncertainties in
supernova (SN) rates vary by a factor of 2-3.
As a simple estimate, let us use the following values: 0.4, 0.2, 0.3,
and 0.1 SN per year for M82, NGC 253, NGC 4945, and M83,
correspondently.  These values are obtained by scaling the mean SN
rate of NGC 253 (0.2 SN per year; \citet{e1998,p2001}) using the far
infra-red luminosity data given in \citep{btr2000}.  Several
investigations (see, for example, references in \citet{btr2000})
showed that the method based on the far infra-red luminosity allows
one to estimate the relative SN rate with a high precision.  Thus, the
main uncertainty is the rate of SN in NGC 253, however this is a
well-studied galaxy, and all estimates of the SN rate given in the
literature are close to 0.2 SN per year.

In comparison with the Galactic SN rate, these galaxies have
significant enhancement (roughly a factor of 12, 6, 9, and 3
correspondently).  It total, the SN rate in the four galaxies is
$\sim30$ times higher than the one in the Milky Way.  We can expect
proportionally higher number of SGRs (and GFs) from them. With the
Galactic rate of $\sim$ 3 flares in 25 years, for BATSE (4.75 years
equivalent of all-sky coverage) we can expect roughly 6-7 GFs from
M82, 3-4 GFs from NGC 253, 5-6 from NGC 4945, and 1-2 GFs from M83 (in
total about 15-20 GFs from four galaxies during the BATSE life cycle).

Could BATSE observe GFs from these galaxies? It is not a simple
question. Surprisingly, we have no reliable estimate of the peak
luminosity of the initial spikes in giant SGR flares. The problem is
that they are so strong that all detectors get severely saturated
during Galactic giant flares.  The situation was slightly better for
the event of March 5, 1979 \citep{m1979, g1979}, as it happened at a
larger distance (in the LMC).  Nevertheless, Venera 11 and Venera 12
detectors were still saturated.  Using the raw count rate detected by
Konus (see Fig.1, lower curve), one gets the maximal energy flux of
$\sim 0.3 \cdot 10^{-3}$erg cm$^{-2}$ s$^{-1}$.  \citet{g1979}
estimate the peak flux to be $1.5 \cdot 10^{-3}$erg cm$^{-2}$
s$^{-1}$.  This estimate corresponds to the luminosity of $0.8\cdot
10^{45}$ erg s$^{-1}$.
The difference between these two values is probably due to the
correction for the dead time.

To estimate the distance from which such event can be observed by
BATSE, we use the spectrum measured by \citet{g1979} (there exist,
however, a different reconstruction of this GF spectrum by
\citet{fkl1996}, see discussion below) and different versions of the
count rate curve (see Fig. 1). The first version is just the raw count
rate and can be considered as a conservative lower limit. It
corresponds to the energy release in the initial spike $2 \cdot
10^{43}$ erg.  The second version is a narrow top spike reaching the
level of count rate $10^6$ cts~s$^{-1}$ corresponding to the peak flux
$1.5 \cdot 10^{-3}$~erg cm$^{-2}$ s$^{-1}$ and $0.45 \cdot
10^{44}$~erg energy release. The third version corresponds to the same
peak intensity but a wider top, and therefore to a larger total energy
release $0.6 \cdot 10^{44}$~erg. The reconstruction of the profile is
somewhat arbitrary (the third version is the closest to the
reconstruction by \citet{g1979}) and should be treated simply as an
illustration of possible variations.

In each case, the spectrum by \citet{g1979} was folded with the BATSE
detector response matrix \citep{petal1999} at a random orientation of
the satellite relative to to the burst arrival direction. Then, the
simulated counts were added to one of the real background fragments
sampled from the BATSE continuous archive records with simulated
Poisson noise in 64 ms bins.  Finally, the BATSE triggering scheme was
applied to each synthetic burst.  The probabilities of BATSE
triggering vs. the distance to the source are given in Fig 2. Curves
in this figure (and in the following one) are normalized in such a way
that the asymptotic value, which is reached at small distances,
represents the sky coverage of detectors.

\begin{figure}
\includegraphics[width=320pt]{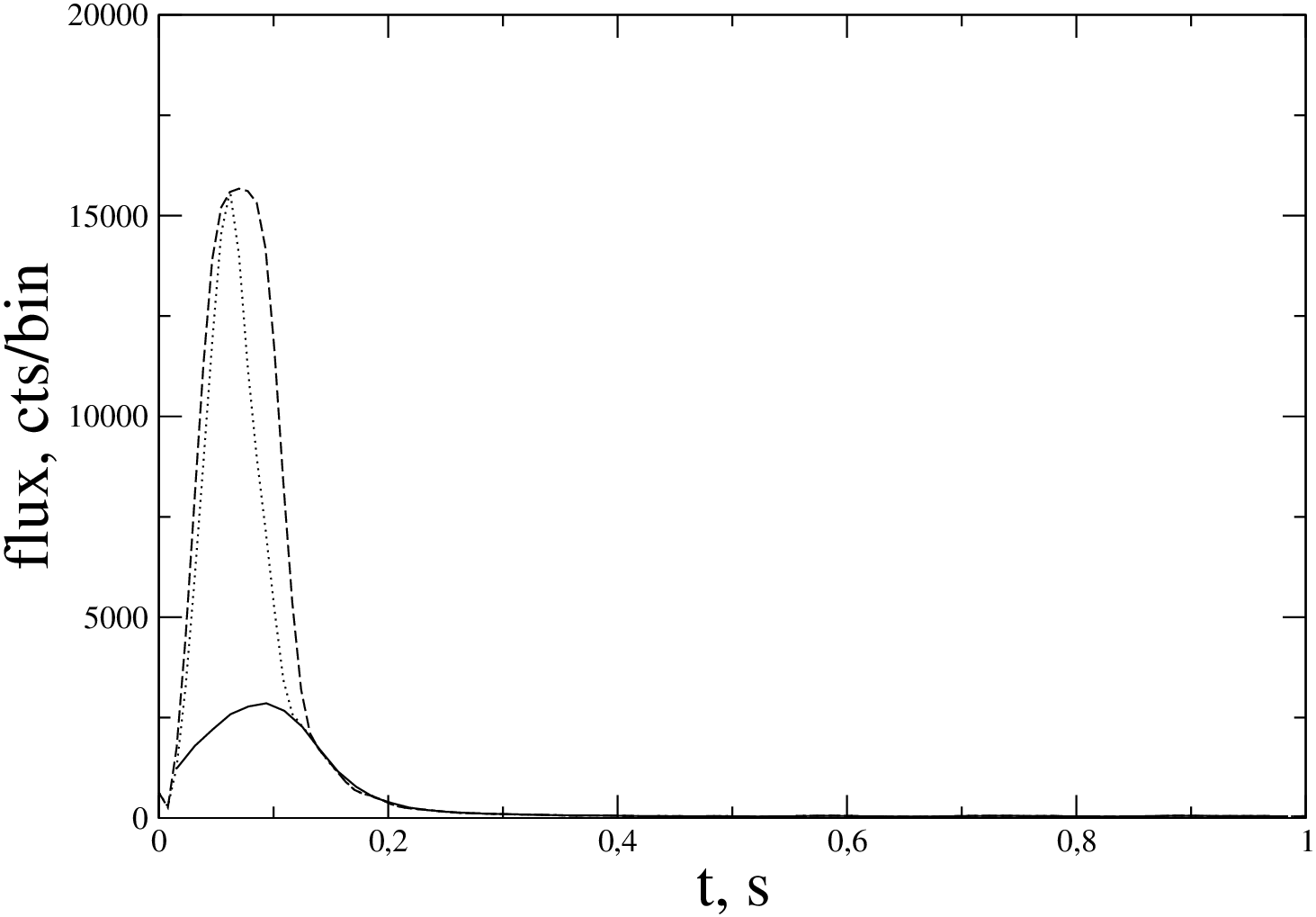}
  \caption{Assumed time profiles of the initial spike of the 5th March 
1979 event. Different versions of the reconstruction are shown. Solid curve: the raw count rate
(subject of saturation), dotted curve: the reconstruction up to 
1.5$\cdot 10^{-3}$ erg cm$^{-2}$ s$^{-1}$ as a narrow top spike.
Dashed curve: the reconstruction to the same level but as a wider top spike.
Curves are smoothed.}
\label{profile}
\end{figure}

In the first case (solid curve, the Konus raw counts), the only large
galaxy in the detectable range is M31. No appropriate candidate for
the GFs from M31 has been detected by BATSE \citep{b2001}.  This is
not surprising as it is not expected that the SGR activity in M31 is
higher than in our Galaxy, and BATSE during its lifetime observed only
one GF -- it is the doubtful event from SGR 1627-41 which is not
considered to be real GF by many authors.
In the second and the third cases (dotted and dashed lines), GFs from
the four near-by galaxies with high SFR mentioned above are
detectable, albeit as fairly weak bursts with a poor angular accuracy.


It is useful to check whether there are potential SGR candidates in
these four galaxies in the BATSE catalogue. We have to look for short
bursts with $T_{90}$ less than 2 seconds at least (the burst from SGR
1627-41 was longer than initial strong spikes from the three other
SGRs). In the duration table of the BATSE catalogue the number of GRBs
shorter than 2 s is 500. The expected number of chance overlaps of
their error boxes with the four galaxies is 9.4 (2.36 per a local
object).  Actually, we have 12 overlaps of 11 GRBs which is consistent
with the expectation for chance coinsidence.  We added a few
overlapping GRBs that are not in the duration table but have
approximate estimate of duration within 2 s.  All these short GRBs are
given in Table 1.
For each burst, we give its trigger number, coordinates, error box
radius, $T_{50}$ and $T_{90}$, energy release in the source at the
distance corresponding to the galaxy with which the error box overlap,
and hardness ratios (counts in BATSE channels 2 (50 - 100 keV) and 3
(100 - 300 keV) respectively to that in channel 1 (25 - 50 keV)).
Coordinates and error box radii are given in degrees.

\begin{figure}
  \includegraphics[width=320pt]{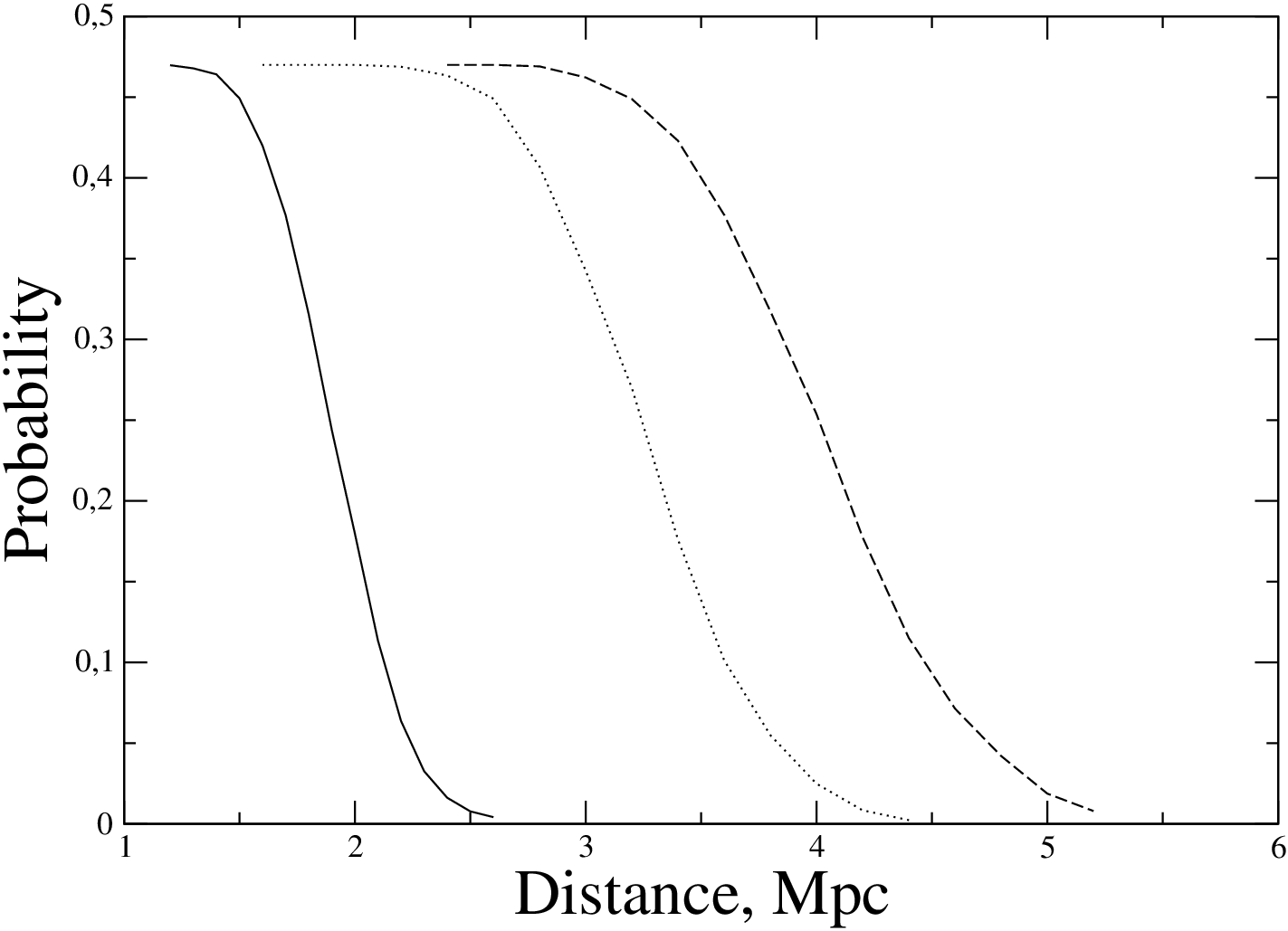}
  \caption{The probability of the BATSE detection of a giant flare 
similar to the 05 March 1979 as a function of distance. The real BATSE exposure factor, 
the representative background sample, the detector response matrix 
and the triggering procedure are taken into
account. Curves correspond to different versions of the
time profiles shown in Fig.1.
For small distances, curves approach the asymptotic value defined by the sky
coverage of BATSE.
}
 \label{eff}
\end{figure}


Can some of these events be the GFs originated in the four galaxies?
Their energy releases are comparable to that estimated for the March
5, 1979 event.

\begin{table*}
 \centering   
 \begin{minipage}{140mm}
  \caption{GRBs coincident with SF galaxies}
  \begin{tabular}{@{}lrrrrrrrr@{}}
  \hline
Trigger  &$\alpha$ & $\delta$ & Error & $T_{50}$,& $T_{90}$, & Energy  &
Ratio 2/1& Ratio 3/1\\
number   &         &          & box   &    s     &    s       & $\times 10^{44}$erg &&\\
&&&&&&& \\
\hline 
&&&&&&& \\
M82 &148.95&69.68&&&&& \\
 2054 & 164.33 & 66.15&  17.91&     ---  &    $\sim 1$ &      --- & ---&\\
 3118 & 117.57 & 80.37 & 23.0 &  0.136  &     0.232 &     1.1 &  3.4&5.0\\
 6255 & 148.68&  60.79 & 12.71 &  ---       &   $\sim 0.4$ &         & 1.2&1.6   \\
 6547 & 155.18&  62.23 & 13.58 &  0.029   &    0.097 &    0.37 & 1.7&2.1\\
 7297 & 140.07 & 76.39 & 9.53   & 0.438  &     1.141  &  2.1  &2.&3.4 \\
 7970 & 136.87 & 64.49&  8.48   & 0.157 &      0.387   &  1.3  &1.1&0.9  \\
&&&&&&& \\
\hline
M83 &204.25&-29.87&&&&& \\
 1510 & 198.84 &-34.35 & 7.29  &   ---     &  $\sim 0.1$  &      --- & 1.3& 1.7 \\
 2384&  203.8 &-18.21  &17.81    & 0.128    &   0.192 &   0.50 & 1.3&1.3\\
 2596&  211.51& -27.07 & 19.74    &---     &  $\sim 0.3$ &   ---   & 2.0 &3.0  \\
 5444  &199.44 &-31.51 & 4.94 &    ---   &  $\sim 0.1$   & ---      &1.6&1.4 \\
 6447 & 191.44 &-36.6  &14.77  &   0.256     &  1.024  &  1.2 & 1.5& 1.9\\
 7361  &204.17 &-28.29 & 7.28    & 0.960   &    1.856   & 1.6 & 1.9 &3.1\\
 7385 & 203.02 &-27.81 & 3.59     &---    &  $\sim 0.2$     & ---   & 1.4 & 1.4\\
 8076&  199.39 &-29.98 & 7.39     &0.075 &      0.218   & 1.4 & 1.9 &7.9\\
&&&&&&& \\
\hline

NGC 253 &11.9&-25.3&&&&& \\
  2312 & 14.72& -33.56&  8.93&     0.112&       0.272 &  0.87 &  1.2 & 12.2\\
  7591 & 15.75& -32.66&  8.03&     ---  &      $\sim 0.5$ & ---      & 0.9&0.9\\
&&&&&&& \\
\hline

NGC 4945 &196.5&-49.5&&&&& \\
 2800 & 200.29& -47.94&  15.92 &     0.320  &     0.448&  1.3 & 1.6&2.1\\
 3895 & 189.39& -47.72&  6.99 &      0.384   &    0.768&  1.3 &$\ga$1.4&$\ga$2.0 \\
 6447 & 191.44& -36.6 & 14.77&       0.256    &   1.024&  1.2 &1.5&1.9\\
&&&&&&& \\
\hline
\end{tabular}
\end{minipage}
\end{table*}  

If we accept the requirement that the time profiles of GFs should be
smooth structureless pulses same as the 05 March 1979 event, then we
have to exclude four events (triggers 3895, 6255, 6547, 7385) from the
list since they have a substructure.  If we require that the duration
of GF spikes is between 0.1 and 0.3 as that of the three detected GFs
then we have to exclude triggers 2054, 7297, 6447, 7361, 3895.  If we
suggest that the spectrum, measured by Golenetskii et al. (1979),
represents a typical spectrum of a GF, then we have to exclude almost
all events.

Indeed, this spectum, once folded with the BATSE detector response
matrix gives the following count ratio in the three energy channels
(1:2:3): 1:1.36:0.58.  All events are much harder except triggers 7970
and 7591 which are just slightly harder.

To what extent should we rely on the spectrum by \citet{g1979}?
\citet{fkl1996} reanalyzed the ISEE-3 data for this event and obtained
much harder spectrum which is inconsistent with the Konus data.  It
should be noted that both reconstructions have their own problems.
The Konus data are integrated over the 3.28~s time interval and are
contaminated by approximately 1/3 of photons from the softer pulsating
tail\footnote{See the raw count rate curve at
  http://www.ioffe.rssi.ru/LEA/SGR/Catalog/Data/0526/790305.htm}.  The
ISEE-3 detector observed the flare through the spacecraft, and the
reconstruction relies on the difficult simulation of the photon
transfer through the instrument with a complicated matter
distribution.

We should recognize that we have no solid hypothesis of the GFs
spectra: the data are available only for one event and are rather
ambiguous.  If we still rely on the Konus spectrum as on the one,
obtained in a more straightforward way, then we have to accept two
events as a conservative upper limit to the observed number of GFs
from the four galaxies. In this case, the 90\% upper limit on the
expected number of observable GFs (i.e. with the energy release $>0.5
\cdot10^{45}$ erg, see Fig. 2 and Table~1) in these galaxies during
BATSE exposure is $\sim$~5~(i.e. $\sim 1$ yr$^{-1}$ per all four
galaxies).  The rate of such GFs in our Galaxy (not per source!)
should be $\sim 30$ times less, or $\sim 1/30$~yr$^{-1}$.  This is
somewhat smaller than has been observed.

If we admit an arbitrary hardness for GFs, then we have 10 candidates
with suitable time profiles and durations and the above constraint
relaxes to $\sim 1/10$ yr$^{-1}$ which is in a good agreement with the
observations.


\section{Hyperflares in the 50 Mpc vicinity}

\begin{figure}
\includegraphics[width=340pt]{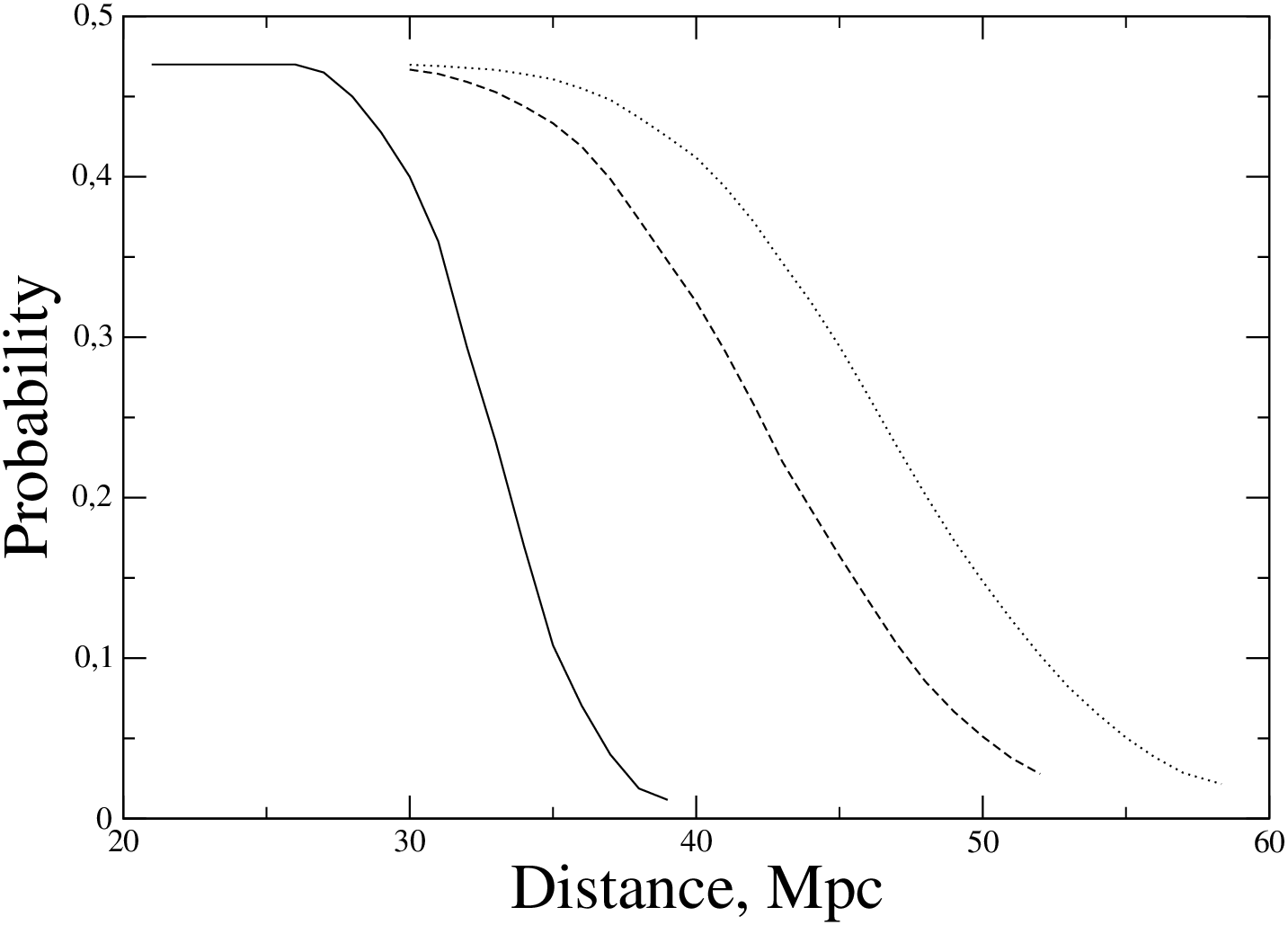}
\caption{The same as Fig. 2, but for a hyperflare with the energy
  release $2\,10^{46}$ erg in the initial spike and the same spectrum
  as for the GF of 05 March 1979.  Dashed and dotted curves correspond
  to different time profiles as shown in Fig. 1. The solid curve
  corresponds to the thermal spectrum with T=200~keV as suggested in
  \citet{h2005}; for this curve the time profile corresponds to the
  dotted curve in Fig.~1.}
 \label{46eff}
\end{figure}

The situation with supergiant flares like that of December 27, 2004 is
quite different since the BATSE sampling volume for such events is
larger by more than 3 orders of magnitude (i.e. accessible distance is
larger by a factor of 10). The data indicate that the spectrum of this
flare is much harder than that of the March 5, 1979 event: according
to \citet{h2005} the spectrum of the initial spike of the December 27,
2004 flare can be described by a blackbody with the temperature of 200
keV. It seems only natural that the events that differ by two orders
of magnitude in the energy release have different spectra.  Fig 3
shows the probability of detection of $2 \cdot 10^{46}$ erg flare by
BATSE for two spectral shapes: as suggested by \citet{h2005} and for
the spectum of the March the 5th event from \citet{g1979}. The
sensitivity is lower in the case of a harder spectrum due to a smaller
number of photons in the 50 - 300 keV band at the same energy release.

The largest structure inside the sampling distance $R \sim 50$ Mpc is
the Virgo galaxy cluster (the cluster center is $\sim 17$ Mpc away,
and the approximate coordinates of the center are $\alpha =
187.5^{\circ}$, $\delta = 12.5^{\circ}$). It contains about 1300
galaxies including 130 spirals (see \citet{bts1987} for details).
Total star formation rate in the cluster is few hundred times larger
than in our Galaxy. BATSE should be able to detect supergiant flares
from the Virgo cluster as fairly strong bursts. We selected short GRBs
detected by BATSE with $0.05 < T_{50} < 0.7$~s. There are 402 such
events.  Only two of them are projected onto the Virgo cluster
(assuming it as a circle with $10^{\circ}$ radius).\footnote{The
  expected number of chance projections is about 3.}  Their trigger
numbers are 2896 (coordinates: $\alpha=180^{\circ},\,
\delta=8.92^{\circ}$, energetics: 1.8$\cdot 10^{46}$ erg at 17 Mpc)
and 6867 (coordinates: $\alpha=185.37^{\circ},\,
\delta=10.02^{\circ}$, energetics: 0.3 $\cdot 10^{46}$ erg).  Three
more events have error circles overlapping with Virgo. This result,
again, is within the expectation for a chance projection.  Again, we
have no evidence of any HF detections, this allows us to put a 90\%
upper limit on the event rate: $\sim 2$ HFs in the Virgo cluster
during the BATSE exposure (assuming 2 detected at 3 expected
coincidences and 2 expected intrinsic).
It implies that on the 27th of December 2004, an exceptionally rare
event has been observed. The rate of such bursts (with energy release
in the initial spike above $\sim 5 \cdot 10^{45}$ erg) is below
$10^{-3} \times SFRV_{500}$~yr$^{-1}$ per galaxy, where $SFRV_{500}$
is the SFR rate in the Virgo cluster divided by 500 galactic SFRs:
$SFRV_{500}$=(SFR in Virgo)/(SFR in the galaxy $\times 500$).  This
constraint coincides with that by \citet{ngpf2005} made with a
different method. When this work was completed in its original form,
the paper by \citet{petal2005} appeared. These authors presented
(without a detailed discussion) a similar constraint, still 3 times
higher, using the Virgo cluster argument.

There are two other promising candidates for the HF detection within
50 Mpc outside the Virgo cluster. These are Arp 299 \citep{n2004} and
NGC 3256 \citep{l2004}, two galaxies with extreme star formation rate
(``supernova factories''). The total star formation rate in these
galaxies is few times lower than that in the Virgo cluster, therefore
they are a less probable source of HFs in the BATSE data.

Nevertheless, these galaxies are of great interest since they are
well-localized and can lead to measurements with a better angular
resolution. A number of candidates for HFs from these galaxies is
given by Popov (private communication\footnote{See the e-print
  astro-ph/0502391.}).  It is interesting to note, that the same two
galaxies were discussed by \citet{sgm2002,gms2003} as possible sources
of ultra high energy cosmic rays. Together with the recent suggestion
by \citet{e2005} it brings another flavour to the problem of high
energy activity of magnetars and its link with star-forming galaxies.



\section{Discussion}


We do not see any convincing BATSE detections of SGR GFs from the
nearby star-forming galaxies.  This non-detection allows us to put a
constraint on the total galactic rate of GFs and HFs with the energy
release in the initial spike $> 0.5 \cdot 10^{44}$~erg. This rate has
to be less than $1/25$ yr$^{-1}$ (note that this estimate is based on
the assumption of a low hardness of GFs, see Sec.~2). The observations
of flares from the sources in our Galaxy indicate that the rate of GFs
+ HFs is higher, still, we can conclude that most of them have energy
release in the initial spike $<0.5 \cdot 10^{44}$ erg. The only
evident exception is the flare detected on the 27th of December of
2004, therefore, this upper limit is not in a conflict with the data.

The absence of detections of hyperflares from the Virgo cluster makes
the recent hyperflare of SGR 1806-20 an exceptionally rare event. A
possible beaming of the emission does not change the conclusion: in
this case we just have to state the same about {\it the observational
  probability} of such event. However, the conclusion is based on the
flare energetics calculated for the 1806-20 distance estimate of 15
kpc. If it is less then 5 kpc, then BATSE could not observe
hyperflares from the Virgo cluster and the constraint should be
relaxed.  However, recent analysis \citep{mg2005} suggests that the
distance is $>$ 6 kpc.

In any case, the conclusion by \citet{h2005} that a large fraction of
short GRBs detected by BATSE can actually be the initial spikes of
extragalactic hyperflares seems too enthusiastic. If the distance
estimate 15 kpc is correct, then the Virgo constraint is valid and we
can renormalize it to the sampling sphere of the radius of 50 Mpc.
The average total star formation rate in this sphere is $\sim$ few
thousand $M_{\sun}$ per year. This estimate can be obtained in several
ways.  For example, \citet{d2001} uses the following expression to
obtain an estimate of a number of galaxies similar to the Milky Way: $
N_{\mathrm{Gal}}=0.0117 \, h_{65}^3 \, R_{\mathrm{Mpc}}^3. $ For
$R=50$ Mpc we obtain about 1500 galaxies. So, for 4.5 years of
observation we can expect nearly 800 GFs and about 200 HFs assuming 3
GFs and 1 HF observed in the Milky Way in 25 years.  Similar estimates
can be obtained using estimates of \citet{bcw2004} and
\citet{gza1995}.  \citet{bcw2004} provide the following value for SFR
density at $z=0.1$: $0.01915\, M_{\odot}/{\rm yr}/{\rm Mpc}^3$. Inside
50 Mpc it gives $\approx 10^4\, M_{\odot}/{\rm yr}/{\rm Mpc}^3$.  SFR
for the Milky way is estimated to be few solar masses per year.  So,
the ratio is about few thousands.  \citet{gza1995} estimate SFR in
star-forming galaxies for $z\la0.045$ as $0.013 \, M_{\odot}/{\rm
  yr}/{\rm Mpc}^3$. It gives $\approx 6800 \, M_{\odot}/{\rm yr}/{\rm
  Mpc}^3$ inside 50 Mpc. All three estimates are in good agreement.
Comparing these values of SFR with few solar masses per year in our
Galaxy one concludes that BATSE could observe $\sim 30 \cdot
SFRV_{500}$ supergiant flares during its 4.75 years of full-sky
exposure, i.e not more than a few percent of the total number of short GRBs. 


In this note, as in the previous literature, we assume the rate and
luminosity of GFs to be constant. However, it should be considered as
only a zeroth approximation, since all types of NS activity usually
decrease with time (for example, rate of glitches, \citet{ah1983}).
If one hypothesizes that the rate of GFs decays with time as $\propto
t^{-\alpha}$, then two interesting consequences can be discussed.  The
first one is the following. For $\alpha>1$ it becomes more probable to
discover a younger magnetar (if energies of flares are the same for
all ages).  In that case one can safely claim that in our Galaxy there
are no magnetars younger than the four known.  Then it is necessary to
note, that for larger $\alpha$, the rate of flares in the magnetar
youth becomes so high, that the energy of the magnetic field, $\sim
10^{47}\, B_{15}^2$~erg, is not sufficient to support numerous GFs
with luminosities similar to the one of March 5, 1979.  This can
explain the fact that no good GF candidates were found from
star-forming galaxies.  In the four near-by star-forming galaxies
there should be SGRs $\sim10$ times younger than the Galactic ones; in
galaxies like Arp 299 and NGC 3256 we expect to find magnetars with
ages about few tens of years. If they produce frequent bursts, then
non-detection should mean that their luminosities are lower than those
exhibited by the galactic sources.

As it is noted by \citet{h2005}, {\it Swift} gives an excellent
opportunity to observe extragalactic GFs and HFs of SGRs. We would
like to emphasize that the most promising targets for such observation
are the Virgo cluster (for HFs) and galaxies M82, M83, NGC 253, and
NGC 4945 (for GFs).\footnote{Long pointings of {\it Integral} in the
  direction of the Virgo cluster potentially also can result in
  detection of GFs or/and HFs. Unfortunately, Integral Galactic plane
  scans do not cover Virgo or any of the six galaxies discussed in
  this paper.}  Of course, due to the large field of view of {\it
  Swift}, several objects can be observed simultaneously.  The
possibility to detect a very strong HF from a young SGR, as discussed
by \citet{h2005}, is much higher in the case of galaxies with extreme
star formation. Arp 299 and NGC 3256 can be good targets for such
observations.
 
\section*{Acknowledgments}
We thank Drs. J. Poutanen and M. Prokhorov for useful discussions, and Dr.
V. Belokurov for comments on the manuscript.  We
appreciate useful comments of the unknown referee of the first version
of this paper.  The work of S.P. was supported by the RFBR grants
04-02-16720 and 03-02-16068, and by the ``Dynasty'' Foundation
(Russia).  The work of B.S. was supported by the RFBR grant
04-02-16987.


\end{document}